\documentclass[12pt]{article}  
\newcommand{\be}{ \begin{equation}}
\newcommand{\ee}{\end{equation}} 
\begin{document} 
\def\theequation{\arabic{section}.\arabic{equation}} 
\begin{titlepage} 
\title{Dark energy, wormholes, and the Big Rip} 
\author{V. Faraoni$^1$ and W. Israel$^{2,3}$ \\ \\ 
{\small \it $^1$ Physics Department, Bishop's University}\\
{\small \it Lennoxville, Quebec, Canada J1M 1Z7}\\\\
{\small \it $^{2}$ Department of Physics and Astronomy, University of 
Victoria}\\ 
{\small \it P.O. Box 3055, Victoria, B.C., Canada V8W~3P6}\\\\
{\small \it $^3$ Canadian Institute for Advanced Research Cosmology 
Program}
}
\date{} \maketitle 
\thispagestyle{empty} 
\vspace*{1truecm} 
\begin{abstract} 
The time evolution of a wormhole in a Friedmann universe approaching the 
Big Rip is studied. The wormhole is modeled by a thin spherical shell 
accreting the superquintessence fluid -- two different models are 
presented. Contrary to recent claims that the 
wormhole overtakes the expansion of the universe and engulfs it before the 
Big Rip is reached, it is found that the wormhole becomes asymptotically 
comoving with the cosmic fluid and the future 
evolution of the universe is fully causal.
 \end{abstract} \vspace*{1truecm} 
\begin{center}  Keywords: wormholes, dark energy, Big Rip
\end{center}
\begin{center} PACS: 04.20.-q, 98.80.-k, 04.20.Jb
\end{center}
\setcounter{page}{0}
\end{titlepage}

\def\theequation{\arabic{section}.\arabic{equation}}


\section{Introduction}
\setcounter{equation}{0}
\setcounter{page}{1}

The 1998 discovery that the expansion of the universe is accelerated 
\cite{SN}, confirmed by the analysis of the cosmic microwave background 
power spectrum \cite{Netterfieldetal,Hinshawetal}, has 
led to postulate, as a possible explanation within the context of general 
relativity, the existence of a form of {\em dark energy} or {\em 
quintessence} with negative pressure $P<-\rho/3$ (where $\rho$ is the dark 
energy density). The dark energy is dynamically irrelevant in the earlier 
stages of the evolution of the universe and starts dominating the cosmic 
dynamics only at recent times.  At present the observational data favour 
an even more exotic dark energy with effective equation of state 
$P<-\rho$ violating the weak energy condition \cite{superacceleration}. 
This violation is rather disturbing because it allows, in principle, for 
exotic 
solutions of general relativity such as wormholes and warp drives and the 
possibility of time travel associated with them. Another concern is that a 
universe dominated by dark energy with effective equation of state 
parameter $w \equiv P/\rho <-1$ may end its existence at a Big Rip 
singularity at which the scale factor and the energy density and pressure 
of quintessence diverge at a finite time in the future \cite{BigRip}. Dark 
energy with $w<-1$ is called {\em superquintessence} or {\em phantom 
energy}.

Recently, the evolution of a wormhole embedded in a 
Friedmann--Lemaitre--Robertson--Walker (hereafter ``FLRW'') universe 
approaching the Big Rip was 
studied by Gonzalez--Diaz \cite{Gonzalez}, with the conclusion that the 
wormhole accreting 
superquintessence expands faster than the background FLRW universe and 
that the radius of the wormhole throat diverges before the Big Rip is 
reached -- thus the wormhole engulfs the entire universe, which will 
re--appear from the other wormhole throat. The resulting spacetime is not 
globally hyperbolic and is acausal with closed timelike curves threading 
the wormhole throat. Such bizarre scenarios are potentially of interest as 
constraints: if it can be established that phantom energy leads in 
principle to unacceptable consequences, this may be sufficient to rule out 
its existence.

The analysis of Ref.~\cite{Gonzalez} is based on a qualitative estimate of 
the accretion rate of phantom energy onto the wormhole 
and the rate of variation of the 
throat 
radius, which borrows from a recent study of a similar problem of 
accretion of dark energy onto black holes \cite{Babichevetal}. The 
purpose of this work is to present exact solution models of a wormhole 
immersed in a spatially flat FLRW universe and to compare the expansion of 
the wormhole throat with that of the cosmic fluid as the Big Rip is 
approached. Two different and rather general wormhole models are studied. 
In the first model it turns out that, in the simplifying approximation of 
stationary accretion advocated in Ref.~\cite{Gonzalez}, the wormhole is 
asymptotically comoving with the FLRW background as the Big Rip is 
approached -- the size of the wormhole with respect to a comoving ruler 
does not increase and  the universe cannot disappear within the wormhole.
This conclusion holds also in the second wormhole model (presented in 
Sec.~5), which is more 
general.


\section{A thin shell wormhole embedded in a spatially flat FLRW universe}
\setcounter{equation}{0}

We study an exact solution of the Einstein equations describing a  
spherically symmetric thin shell wormhole 
embedded in a  spatially flat FLRW universe described by the metric 
\cite{footnote1}
\be \label{metric}
ds^2=-dt^2 +a^2(t) \left[ dr^2 +r^2\left( d\theta^2+\sin^2 \theta 
\, d\varphi^2 \right) \right] 
\ee
in comoving polar coordinates $\left( t, r, \theta, \varphi \right)$.
The thin shell is located on a surface $\Sigma$ of constant comoving 
radius. Two  FLRW regions joining smoothly on the surface $\Sigma$ 
constitute the regions ``above'' and ``below'' $\Sigma$. The assumption 
that the line element 
is given by 
eq.~(\ref{metric}) implies that the wormhole shell does not perturb the 
surrounding universe (see the discussion in Sec.~4). The equation of the 
shell $\Sigma$ is
\be \label{shellequation}
r = \frac{ R(t) }{a(t)} \equiv \mbox{e}^{-\alpha(t)} \;,
\ee
where $R(t)$ is the comoving radius of the shell and the function 
$\alpha(t)$ is introduced for later convenience. The normal vector to the 
shell, obtained by differentiating 
eq.~(\ref{shellequation}), is 
\be
N^{\alpha} = \left( -\alpha_t \, \mbox{e}^{-\alpha}, a^{-2}, 0, 0 \right)
\ee
where a subscript $t$ denotes differentiation with respect to the 
comoving time $t$ of the FLRW background. The norm squared of $N^{\alpha} 
$ is 
\be
g_{\alpha\beta}N^{\alpha} N^{\beta} =-\alpha_t^2 \, 
\mbox{e}^{-2\alpha} + \frac{1}{a^2} 
=\left( \frac{\beta}{a} \right)^2 \;,
\ee
where 
\be \label{beta}
\beta \equiv \sqrt{ 1-\alpha_t^2 a^2 \mbox{e}^{-2\alpha} } \equiv 
\sqrt{1-v^2} 
\ee
can be interpreted  as the inverse of the Lorentz factor constructed with  
the velocity of the shell relative to the background
\be \label{v}
v \equiv - \alpha_t \, R \;.
\ee

By normalizing $N^{\alpha} $ one obtains the unit normal to the surface 
$\Sigma$ 
\be \label{shellnormal}
n^{\mu}=\left( \frac{ -\alpha_t R}{\beta}, \frac{1}{a \beta} , 0,0 \right) 
\ee
and the three-metric on the surface $\Sigma$ is given by
\be \label{3metric}
\left. ds^2 \right|_{\Sigma} = -\beta^2 dt^2 +R^2(t) \left( 
d\theta^2+\sin^2\theta \, d\varphi^2 \right)=
-d\tau^2 +R^2(\tau) d\Omega^2 
\ee
in coordinates $\left( t, \theta, \varphi \right) $ on $\Sigma$, 
where $\tau$ is the proper time in the shell's frame (defined by $d\tau 
=\beta dt$) and $d\Omega^2$ is the line element on the unit two--sphere.  
By 
using the tetrad
\be
\left\{ 
e^{\alpha}_{(t)} , e^{\alpha}_{( \theta )} ,  e^{\alpha}_{( \varphi)} \right\}
= \left\{  
\beta \, \delta^{\alpha}_{ t}  , 
\delta^{\alpha}_{ \theta } ,  \delta^{\alpha}_{ \varphi}  \right\}
\ee
($ a,b=t, \theta, \varphi $), one computes the extrinsic curvature of 
$\Sigma$
\be \label{extrinsic}
K_{\alpha\beta}=e_{\alpha}^{(a)} \, e_{\beta}^{(b)}\, \nabla_a n_b =
 e_{\alpha}^{(a)}\, e_{\beta}^{(b)} \left( \partial_a n_b 
-\Gamma^c_{ab}n_c \right) \;.
\ee
Its components are found to be
\begin{eqnarray}
{K^t}_t & =&  -\frac{1}{\beta^2} \, \partial_t \left( \frac{ \alpha_t \,
R}{\beta} \right) \;, \\
&& \\
{K^{\theta}}_{\theta} & =& 
{K^{\varphi}}_{\varphi}= \frac{\beta}{ R}  - \frac{ \alpha_t \, 
R_t}{\beta} 
\end{eqnarray}
and its trace  is
\be \label{kappa}
K={K^t}_t + {K^{\theta}}_{\theta}+ {K^{\varphi}}_{\varphi}=
- \, \frac{1}{\beta^2} \, \partial_t \left( \frac{\alpha_t \, R}{\beta} 
\right) +\frac{2\beta}{R} -\frac{2\alpha_t \, R_t}{\beta} \;.
\ee
Since the shell has no interior and two FLRW regions match on $\Sigma$, 
the jumps of $ {K^a}_{b} $ and of $K$ at $\Sigma$, which appear in 
the Einstein equations at the shell, are given by
\be
\left[ {K^a}_b \right]= 2 {K^{a}}_{b} \;, \;\;\;\;\;\;\;\;\;\;
\left[ K \right]= 2 K 
\ee

The Einstein equations at $\Sigma$ are \cite{BarrabesIsrael}
\be  \label{efeatsigma}
\left[ {K^a}_b - K {\delta^a}_b \right] =-8\pi {S^a}_b \;,
\ee
 where
\be 
S_{ab}= \left( \sigma+P_{(\Sigma)} \right) u^{(\Sigma)}_a u^{(\Sigma)}_b + 
P_{(\Sigma)} \, 
g_{ab}\left.\right|_{\Sigma}
\ee
is the stress-energy tensor of the (exotic) matter living on the shell 
and $u^{\mu}_{(\Sigma )} $ is the four-velocity of the shell. 
$\sigma$ and $P_{(\Sigma)}$ are, respectively, the (surface) energy 
density 
and 
the pressure on the shell. The time-time component of the Einstein 
equations 
yields
\be \label{sigma}
\sigma= -\frac{1}{2\pi} \left( \frac{\beta}{R} -\frac{\alpha_t R_t}{\beta} 
\right) \;,
\ee
while the $\theta\theta$ or the $\varphi\varphi$ component yields
\be \label{pesse}
P_{(\Sigma)}= \frac{1}{4\pi} \left[ -\, \frac{1}{\beta^2} \partial_t \left( 
\frac{ 
\alpha_t R}{\beta}\right) +\frac{\beta}{R} -\frac{\alpha_t R_t}{\beta} 
\right]
=-\frac{\sigma}{2} - \frac{\partial_t \left(u^{\alpha} n_{\alpha} 
\right)}{4\pi \beta^2}  \;,
\ee
where $u^{\mu}$ is the four-velocity field of the cosmic fluid. The 
effective equation of state of the exotic matter on the shell is 
given by
\be
\sigma+2P_{(\Sigma)}= -\, \frac{1}{2\pi \beta^2} \, \partial_t \left( 
\frac{\alpha_t 
R}{\beta} \right) = -\, \frac{1}{2\pi \beta^2} \, \partial_t \left( 
u^{\alpha}n_{\alpha} \right) \;,
\ee
while the material energy residing on the shell is
\be \label{M}
M \equiv 4\pi R^2\, \sigma =-2\left( \beta R -\frac{\alpha_t R_t 
R^2}{\beta} 
\right) \;.
\ee
The relative speed $ v= - \alpha_t \, a \, \mbox{e}^{-\alpha} $ between  
the shell and the background fluid, $\beta=\sqrt{1-v^2} $, and $ 
u^{\alpha}n_{\alpha}= - v/\beta$, are constant when the 
matter on the shell satisfies the equation of state $P_{(\Sigma)} 
=-\sigma/2$.

The Einstein equation
\be
\tilde{K}_{ab} \,  S^{ab}=\left[ T_{\alpha\beta} \, n^{\alpha} n^{\beta} 
\right] \;,
\ee
where 
\be 
\tilde{K}_{ab}=\frac{ K_{ab}^{(+)}+ K_{ab}^{(-)} }{2} =0
\ee
is the mean of the extrinsic curvature on both sides of the shell, 
provides 
the rather obvious matching condition for the energy 
density and pressure of the FLRW cosmic fluid on both sides of $ \Sigma$:
\be
\rho^{(+)} = \rho^{(-)} \;, \;\;\;\;\;\;\;\;
P^{(+)} = P^{(-)} \; .
\ee

It is sometimes convenient to use the parameter $\chi$ defined by
\be
v= - \alpha_t R = - \dot{\alpha}\beta R =\tanh \chi \;,
\ee
or by $ \cosh \chi = \beta^{-1} $, where an overdot denotes 
differentiation 
with respect to the comoving time of the shell $ \tau $ and $ 
\dot{f}=f_t/\beta $ for any differentiable function $f$. 

The four-velocity of the shell 
\be
u^{\alpha}_{(\Sigma )}=\frac{dx^{\alpha}}{d\tau}
=\frac{1}{\beta} \, \frac{ dx^{\alpha}}{dt}=
\frac{1}{\beta} \left( 1, -\alpha_t \, \mbox{e}^{-\alpha} , 0, 0 \right)
\ee
has the properties
\begin{eqnarray}
&& u^{\alpha}_{(\Sigma)} \, n_{\alpha} =0 \;, \\
&& \\
&& u^{\alpha}_{(\Sigma )} \, u_{\alpha}^{(\Sigma )} = -1  \;. 
\end{eqnarray}
The conservation equation (eq.~(5) of Ref.~\cite{BarrabesIsrael}) 
projected along the four-velocity of the shell $u^a_{(\Sigma )} $ yields
\be\label{conservation}
{{S_a}^b}_{;b} \, u_{(\Sigma )}^a= -\left[ 
u_{(\Sigma )}^{\alpha} \, 
T_{\alpha}^{\beta} n_{\beta} \right] \;.
\ee
The left hand side 
$ {{S_a}^b}_{;b} \, u_{(\Sigma )}^a = \left( {S_a}^b u_{(\Sigma )}^a 
\right)_{;b}-
S_a^b \, { u_{(\Sigma)}^a}_{;b} $ of eq.~(\ref{conservation}) reduces to
$  -\left( \sigma u^b_{(\Sigma )} \right)_{;b}-P_{(\Sigma)} 
{ u_{(\Sigma)}^b }_{;b} $, where $\left( \sigma u_{(\Sigma )}^b 
\right)_{;b}= \dot{M} / {\cal A} $, ~~$ {\cal A} \equiv 4\pi R^2(t)$ is 
the (proper) 
surface area of the shell, and $M \equiv  \sigma {\cal A} $ is the 
material energy 
located on 
the shell. Similarly, it is found that $ { u_{(\Sigma )} ^b}_{; b} 
=\dot{ {\cal A}}/ {\cal A}$. 
The stress-energy tensor of the cosmic perfect fluid in the FLRW 
background is
\be 
T_{\alpha\beta}= \left( P+\rho \right) u_{\alpha} u_{\beta} +P 
g_{\alpha\beta} \;,
\ee
where $u^{\alpha} $ is the four-velocity of the FLRW comoving observers. 
Then
\be
u^{(\Sigma )}_{\alpha} \, T^{\alpha}_{\beta} n^{\beta}=\left( P+\rho 
\right) 
\left( 
u^{\alpha}_{(\Sigma )} u_{\alpha} \right) 
\left( u^{\beta}n_{\beta} \right)+P u_{(\Sigma )}^{\alpha} \, n_{\alpha} 
\;.
\ee
By using the relations $ u_{(\Sigma )}^{\alpha} u_{\alpha}=-\beta^{-1}$, 
$ u^{\beta}n_{\beta} =\dot{\alpha}R$, and $ 
u^{\alpha}_{(\Sigma )}n_{\alpha}=0$ 
one obtains
\be
u^{(\Sigma )}_{\alpha} T^{\alpha}_{\beta} n_{\beta} =-\frac{\left( P+\rho 
\right)}{\beta} \, \dot{\alpha}R \;.
\ee
Since the unit normal $n^{\mu} $ to $\Sigma$ has opposite sign ``above'' 
and ``below'' $\Sigma$, the jump in this quantity is
\be
\left[ u^{(\Sigma )}_{\alpha} \, T^{\alpha}_{\beta} n^{\beta} \right] 
=-\, \frac{2}{\beta}  \left( P+\rho \right)  \, \dot{\alpha} R 
\ee
and the conservation equation (\ref{conservation}) yields, using 
$\dot{\alpha}R= - v/\beta$,
\be \label{accretion}
\dot{M}+P_{(\Sigma)} \dot{{\cal A}}=
\frac{2}{\beta^2} \left( P+\rho 
\right) {\cal A} \, v 
\;.
\ee
This equation describes the rate of accretion of the cosmic fluid by the 
wormhole and can be interpreted as follows. The quantity 
$ \left( P+\rho \right) v $ on the right hand side is the flux 
density of the cosmic fluid crossing the shell $\Sigma$ radially, the factor 2 
arises because the outflow is from both faces of the shell, one factor 
$\beta^{-1}$ comes from the relativistic mass dilation, while another 
factor $\beta^{-1}$ comes from Lorentz contraction in the radial direction 
due to the relative motion of the shell and the FLRW background. Note that 
in a de Sitter background enjoying the equation of state $P=-\rho$ there  
is no accretion on the shell and static solutions with both $\dot{M}$ and 
$\dot{R}$ vanishing (such as those considered in 
Ref.~\cite{Lemosetal} or their generalizations) become possible.

If the strong energy condition is satisfied by the cosmic fluid and 
$P+\rho >0$ then a wormhole shell expanding relative to the cosmic 
substratum ($v > 0$) accretes positive cosmic fluid energy while a shell 
contracting ($ v < 0$) relative to the cosmic substratum experiences an 
energy outflow through  $\Sigma $.

\section{$v=$constant solutions}
\setcounter{equation}{0}

In the special case in which the relative radial velocity of the cosmic 
fluid and the wormhole shell is constant,
\be \label{constantv}
v= - \alpha_t\, R = \tanh \chi=v_0
\;,
\ee
the equation of state of the exotic matter on the shell is 
$P_{(\Sigma)}=-\sigma/2$ 
and one can eliminate $ R(t)$ betwen eqs.~(\ref{constantv}) and 
(\ref{shellequation}) obtaining
\be \label{shellmotion}
- a(t) \, \partial_t\left( \mbox{e}^{-\alpha(t)} \right) +v_0 =0 \;.
\ee
We are interested in a FLRW universe approaching the Big Rip and 
therefore, 
as done in Ref.~\cite{Gonzalez}, we assume for simplicity that the 
equation of state of the cosmic fluid is constant, $P=w\rho$ with $w<-1$. 
This guarantees the occurrence of a Big Rip described by the form of the 
scale factor 
\be  \label{bigrip}
a(t)=a_0 \left( t_{rip} -t \right)^{\frac{2}{3\left( w+1 \right)}} \;,
\ee
which diverges  together with the energy density 
$\rho=\rho_0 \, a^{3 \left( w+1 \right)} $ and the pressure $P=w\rho$ 
as $t\rightarrow t_{rip}$. Under this assumption the constant $v$ solution 
of eq.~(\ref{shellmotion}) for the motion of the wormhole shell is given 
by
\be
\mbox{e}^{ - \alpha(t)} =C v_0 \left( t_{rip}-t \right)^{ 
\frac{3w+1}{3\left( w+1 \right)}} + 
\mbox{e}^{ - \alpha_0}\;,
\ee
where $C$ and $\alpha_0$ are constants. The comoving radius of the shell 
is
\be
R(t)=a(t) \, \mbox{e}^{-\alpha(t) } \simeq a(t) \,  
\mbox{e}^{-\alpha_0} \;:
\ee
it scales asymptotically like the scale factor $a(t)$ and 
hence the wormhole does not overtake the expansion of the universe, 
contrary to what is suggested in Ref.~\cite{Gonzalez}.

The fact that the shell cannot expand faster than the universe 
indefinitely can also be seen by differentiating 
eq.~(\ref{shellequation}) with respect to $t$, which yields
\be  \label{dtshellradius}
\frac{R_t}{R}=H + \frac{v}{R} \;,
\ee
where $H \equiv a_t/a$ is the Hubble parameter of the FLRW universe. When 
$v$ is constant and the radius of the shell expands to infinity the term 
$v/R$ in eq.~(\ref{dtshellradius}) becomes negligible and the expansion 
rate of the shell (with respect to comoving time) coincides with the 
expansion rate of the background universe, even if the shell starts out 
expanding at a faster rate than the background. The final state is 
indistinguishable from one with $v=0$ and it is the 
same irrespective of the initial conditions. (Strictly speaking, this 
argument does not assume 
that the Big Rip is approached, but only that the universe 
expands forever and $v=$const., which in turn implies 
that the wormhole shell expands to infinity, $R(t) \rightarrow +\infty$).

Perhaps a more physical way of looking at this aspect is the following: if 
the shell were to expand faster than the cosmological background in the 
approach to the Big Rip, and given that $H \rightarrow +\infty$ near the 
Big 
Rip, the relative speed $ v$ between the shell and the cosmic fluid would 
increase its magnitude without bound. This would contradict the fact that 
$\left| v \right|$ must be bound by unity, or else the wormhole shell 
becomes tachyonic. Hence the asymptote $ \left| v \right| \simeq 1 $ must 
be approached before the Big Rip is reached, with $\partial_t v 
\rightarrow 0$. A comparison of these results with those 
of Ref.~\cite{Gonzalez} is given in Sec.~6.

\section{Gravitating mass of the wormhole }
\setcounter{equation}{0}

We briefly discuss our assumption that the wormhole 
considered here does not perturb the FLRW surroundings.  This assumption 
must correspond to a zero gravitational mass for the wormhole. In the 
static limit $a\equiv 1$ the mass-energy on the shell is the only 
mass-energy in the entire spacetime. Because the latter is asymptotically 
flat,  the total gravitational mass is given by the Tolman mass 
\cite{Tolman}; for a shell with surface stress--energy tensor $S_{ab}$ in 
an asymptotically flat spacetime the Tolman mass is given by the 
expression \cite{deLaCruzIsrael}
\begin{eqnarray}
M_T & = &  \int_{\Sigma} d\Sigma \, \sqrt{-g_{00}} \left( 
-S^0_0 + S^2_2 + S^3_3 \right) \nonumber \\
&& \nonumber \\
&=& \int_0^{2\pi}d\varphi \int_0^{\pi}d\theta \, R^2 \beta \sin\theta 
\left( 
\sigma+2P_{(\Sigma)} \right)
 = - \, \frac{2R^2 }{\beta} \, \partial_t\left( \frac{\alpha_t\, R}{\beta} 
\right) \;.   \label{Tolman}
\end{eqnarray}
If $\alpha_t R$ is constant, corresponding to constant $v$ and $\beta$, 
the Tolman mass $M_T$ vanishes identically and the wormhole does not 
affect the 
Minkowski background both above and below $\Sigma$. Note that not only the 
material mass $4\pi R^2 
\sigma$, but also the pressure $P_{(\Sigma)}$ contributes to the Tolman 
mass, and 
that the pressure is adjusted to compensate the contribution of the 
material mass in such a way that the Minkowski  background is not 
altered.

For  a wormhole embedded in a FLRW space the Tolman mass is replaced by 
the Hawking quasi-local mass \cite{Hawking,Hayward}: since the metric 
(\ref{metric}) outside the wormhole shell is exactly a FLRW one, the 
quasi-local mass reduces to that of a FLRW space, $E=4\pi r^3 \rho/3$, 
with no contribution from the shell. One concludes that the gravitational 
mass of the wormhole is zero. If instead a spherical black hole is 
embedded in a FLRW background, as described e.g., by the McVittie solution 
\cite{McVittie}, the 
quasi-local mass coincides with the black hole mass times the 
scale factor $a(t)$ \cite{Nolan}.
As an astrophysical object, a  zero-mass wormhole construct located in a 
galaxy would not perturb the orbits of stars that do not fall directly 
into its throat, and it would not cause gravitational lensing (lensing by 
negative mass wormholes is studied in Refs.~\cite{wormholelensing}).


\section{More general wormholes}
\setcounter{equation}{0}

The wormhole model considered in sections~3 and~4 suffers from the 
intrinsic limitation that the wormhole shell is adjusted so that it does  
not 
perturb the surrounding 
cosmological background. It is interesting to ask whether removing this 
assumption has an effect on the results of Sec.~3 about the relative 
motion of the shell and the background  fluid as the Big Rip 
singularity is approached. To this purpose, consider the McVittie metric 
\cite{McVittie}
\be \label{McVittie}
ds^2=-\left[ \frac{B(t,r)}{A(t,r)} \right]^2 \, dt^2 +a^2(t) A^4\left( t, 
r \right)  \left( dr^2+r^2 d\Omega^2  \right) 
\ee
in isotropic coordinates $\left( t,r, \theta,\varphi \right)$, where
\be \label{AB}
A\left(t,r \right) \equiv  1 + \frac{ m(t) }{2r} \;, \;\;\;\;\;\;\;\;\;\;
B\left(t,r \right) \equiv 1-\frac{ m(t) }{2r} \;.
\ee
The modification of the FLRW metric introduced by a non--vanishing 
function $m(t)$ is caused by a spherical wormhole shell $\Sigma$ 
located at the radius
\be \label{rsigma}
r=r_{ \Sigma }( t )=
\frac{R(t)}{ a(t) \, A^2 \left( t,r_{\Sigma} \right) }\;.
\ee
The three--dimensional metric on $\Sigma$ is given by 
\be  \label{3metric2}
{ds^2}\left.\right|_{\Sigma}= 
-\left( \frac{B}{A} \, \mbox{sech} \chi \right)^2  dt^2 
+ R^2(t) \, d\Omega^2     
\ee
in coordinates $ \left( t,\theta,\varphi \right) $ on $\Sigma $ and with 
the function $\chi$ defined by
\be
\tanh \chi(t) \equiv a(t) \, \frac{ A^3 \left( t,r_{\Sigma} \right) }{ B 
\left( t, r_{\Sigma} \right) } \, \frac{dr_{\Sigma} }{dt} \;.
\ee
The unit normal to $\Sigma$ is 
\be \label{unitnormal}
n_{\mu}=\lambda \left( -\,\frac{dr_{\Sigma}}{dt}, 1, 0,0 \right) \;, 
\;\;\;\;\;\;\;\;\;\;  \lambda= aA^2\cosh \chi \;,
\ee
the four-velocity of the shell is
\be 
u^{\mu}_{(\Sigma )}=\frac{A}{B}\, \cosh \chi \left( 1, 
\frac{dr_{\Sigma}}{dt}, 0,0 \right) \;,
\ee
and the four-velocity of the cosmic fluid is
\be
u^{\mu}=\left( \frac{A}{B} ,0,0 ,0 \right) \;.
\ee
The projection of the shell four--velocity on the cosmic fluid 
four--velocity is 
\be
u^{\alpha}_{(\Sigma)}  \, u_{\alpha}=-\cosh\chi= - \, \frac{1}{\sqrt{1-v^2}} \;,
\ee
where the three-dimensional velocity of the shell relative to the cosmic 
fluid is defined as
\be \label{v2}
v\equiv \tanh \chi(t) \;.
\ee
By differentiating $ \left(R/a \right) $ and using the expressions 
(\ref{AB}) of 
$A(t,r)$ and $B(t,r)$ one obtains an equation regulating the dynamics of 
the wormhole shell:
\be   \label{rdotsigma}
\frac{dr_{\Sigma}}{dt}=\frac{1}{AB} \left[ \frac{d}{dt}\left( \frac{R}{a} 
\right)-A \, \frac{dm}{dt} \right] \;.
\ee
The extrinsic curvature of the shell $\Sigma$ given by eq.~(\ref{extrinsic}) has the 
only nonzero components
\be 
{K^t}_t= \lambda \left[ \frac{A^2}{B^2} \left( \frac{d^2 r_{\Sigma}}{ 
dt^2} + \frac{1}{AB \, r_{\Sigma} } \frac{dr_{\Sigma}}{dt} \, \frac{dm}{dt} 
\right)+\frac{m}{a^2A^5B \, r^2_{\Sigma}} \right] \;,
\ee
\be 
{K^{\theta}}_{\theta}={K^{\varphi}}_{\varphi}=
\lambda \left[ \frac{B}{a^2A^5 \, r_{\Sigma}} +
\frac{A^2}{B^2} \frac{d r_{\Sigma}}{dt}
\left( H+\frac{1}{A \,  r_{\Sigma}} 
\frac{dm}{dt} \right) \right] 
\ee
where, again, $H \equiv a_t/a $. The Einstein equations  
(\ref{efeatsigma}) at 
$\Sigma$ then yield the surface density $\sigma $ and the 
pressure $P_{(\Sigma)} $ of the shell material
\be
\sigma=-\frac{ {K^{\theta} }_{\theta} }{2\pi}=-\frac{1}{2\pi} \left[
\frac{B}{a A^3 \, r_{\Sigma} } \cosh \chi +\frac{A}{B}\left( 
H+\frac{1}{A \, r_{\Sigma} } \frac{dm}{dt} \right) \sinh\chi \right] \;,
\ee
\be
\sigma +2P_{(\Sigma)} =  \frac{ {K^t }_t }{2\pi}
= \frac{\lambda}{2\pi} \left[
\frac{A^2}{B^2} \left( \frac{ d^2 r_{\Sigma} }{dt^2} +
\frac{1}{AB r_{\Sigma}} \frac{dr_{\Sigma} }{dt} \frac{dm}{dt} 
\right)+ \frac{m}{a^2 r^2_{\Sigma} A^5 B} \right] 
\ee
and the mass of exotic matter on the wormhole shell can be written, 
using eq.~(\ref{rsigma}),
\be  \label{label}
M\equiv 4\pi R^2(t) \sigma = -2R\left[
\frac{B}{A} \cosh\chi+\frac{AR}{B} \left( H+\frac{1}{A \, r_{\Sigma}} 
\frac{dm}{dt} \right) \sinh\chi \right] \;.
\ee
The  cosmic imperfect fluid is described by the stress-energy tensor
\be
T_{\alpha\beta}=\left( P+\rho \right) u_{\alpha}u_{\beta} 
+Pg_{\alpha\beta} +q_{\alpha}u_{\beta}+q_{\beta}u_{\alpha} \;,
\ee
where the purely spatial vector $q^{\mu} $ (with $ q^{\mu}u_{\mu}=0$) is 
the 
radial energy flux density. The covariant conservation equation of 
mass-energy for the shell, eq.~(\ref{conservation}), yields
\begin{eqnarray}
&& \frac{1}{{\cal A} } \left( \frac{dM}{d\tau_{ \, \Sigma}}+P_{(\Sigma)} \,
\frac{ d{\cal A} }{d\tau_{ \,\Sigma}} \right) =2\lambda\cosh\chi \left\{ 
\left( P+\rho \right) \frac{A}{B} \frac{dr_{\Sigma}}{dt}-q^r \left[ 
\frac{A^2}{B^2} \left( \frac{dr_{\Sigma}}{dt} \right)^2 a^2A^4 +1 \right] 
\right\} \;, \nonumber \\
&& 
\end{eqnarray}
where $ {\cal A} \equiv 4\pi R^2(t)$ and $\tau_{\,\Sigma} $ is the proper 
time of 
the shell defined by the three-metric (\ref{3metric2}) as 
\be
d\tau_{\,\Sigma}= \frac{B}{A} \, \mbox{sech} \chi \, dt \;.
\ee
The Einstein 
equation $ G^1_0=8\pi T^1_0 =8\pi q^r \, u_t  $
determines the radial flux density $q^r$ as
\be
\frac{2m}{AB \, r_S^2} \left( 
H+\frac{1}{m}\frac{dm}{dt} \right) =8\pi q^r \, u_t \;,
\ee
where $r_S(t,r) \equiv a(t)A^2(t,r ) r$. Further use of 
eqs.~(\ref{unitnormal}) and  (\ref{rdotsigma}) yields
\be  \label{accretion2}
\frac{dM}{dt }+P_{(\Sigma)} \, 
\frac{d {\cal A} }{dt} 
=\frac{2}{\beta} \left[ \frac{B}{A} \left( P+\rho \right) {\cal A} v  
+ \frac{A}{B} \frac{ \left( 1+v^2 \right) }{\beta}  
\, \frac{d\left(ma \right)}{dt }\right] \;,
\ee
or
\be \label{accretion3}  
\frac{dM}{d\tau_{ \, \Sigma}}+P_{(\Sigma)} \, 
\frac{d {\cal A} }{d\tau_{ \,\Sigma}} 
=\frac{2}{\beta }  \left[ \left( P+\rho \right)  
{\cal A} \, \frac{ v}{\beta} +  \frac{A}{B} \left( 1+v^2 \right)  
\, \frac{d\left( ma \right)}{d\tau_{ \, \Sigma} }\right] \;,
\ee
in terms of $\tau_{ \, \Sigma}$,
where $\beta =\sqrt{1-v^2} $ again. 
Eq.~(\ref{accretion3}) 
reduces to eq.~(\ref{accretion}) when $m(t) \equiv 0$. The second term on 
the right 
hand side of eq.~(\ref{accretion2}) 
is present even when $v=0$ and describes a contribution to accretion onto 
the shell due to the radial energy flux of cosmic 
fluid -- this term vanishes if 
\be
\frac{1}{m}\frac{dm}{dt}=-H \;,
\ee
equivalent to $m(t)=\mbox{const.}/a(t)$, or $q^r=0$ and $G^1_0=0$. The 
quantity  $m(t) \, a(t) $ appearing in the second term on the right hand 
side 
of eq.~(\ref{accretion2}) coincides with the Hawking quasi-local mass 
\cite{lastfootnote}.   The ``Schwarzschild mass function'' $m_S \left( t, r \right)$ 
instead is defined by 
\be
1-\frac{2m_S \left(t,r \right)}{r_S \left(t,r \right)}\equiv g^{\alpha\beta} 
\left( \nabla_{\alpha} r_S \right)
\left( \nabla_{\beta} r_S \right) \;.
\ee
The definitions of $A,B$, and $r_S$ and the time-time component of the Einstein equations
\be
3\left( \frac{A}{B} \right)^2 \left( H+\frac{1}{rA}\,\frac{dm}{dt} \right)^2=8\pi \rho
\ee
yield
\be
m_S \left(t,r \right)= m(t) \, a(t) +\frac{4\pi}{3} \, r_S^3 \, \rho \;,
\ee
which reduces to the usual Schwarzschild mass in the absence of cosmic fluid.

We are left with the problem of solving for the dynamics of the wormhole 
shell, which is determined by eq.~(\ref{rdotsigma}). The latter can be 
written as
\be  \label{shelldynamics2}
\frac{d}{dt}\left( \frac{R}{a} \right) -A\, \frac{dm}{dt} =\frac{v}{a} 
\left( \frac{B}{A} \right)^2 \;.
\ee
During the approach to a Big Rip singularity in which the scale factor 
$a(t) $ has the form (\ref{bigrip}), the metric component $g_{00}=-\left( 
B/A  \right)^2 $ stays finite even if $m(t)$ diverges. It seems reasonable 
to 
require that both $A$ and $B$ be finite and that the divergence be 
contained in $a(t)$, keeping in mind that the mass of the wormhole shell (which is 
allowed to diverge)  
is not $m(t)$, but is obtained by integrating eq.~(\ref{accretion2}). On 
the other hand, if $v$ is finite ($v\leq 1$) eq.~(\ref{shelldynamics2}) 
reduces to the asymptotic equation
\be
\frac{d}{dt}\left( \frac{R}{a} \right) -A\, \frac{dm}{dt} =0 
\ee

By using the implicit definition (\ref{rsigma}) of $r_{\Sigma}$ one obtains $ B\left( t, 
r_{\Sigma} \right)dr_{\Sigma}/dt=0$, which has the constant solution $r_{\Sigma}=C$, or 
$R(t)=a(t)\left[ C+m(t) +\frac{m^2(t)}{4C}  \right] $. Again, approaching the Big Rip, 
the wormhole shell becomes comoving with the cosmic substratum.


\section{Discussion and conclusions}
\setcounter{equation}{0}

We are now ready to discuss the difference between our results and those 
of Gonzalez-Diaz \cite{Gonzalez}.  Gonzalez-Diaz 
begins by 
considering  a static wormhole solution appearing in 
the appendix of Ref.~\cite{MorrisThorne} (eqs.~(A.28) of 
\cite{MorrisThorne}) and consisting of a spherical thin shell of exotic 
matter with throat radius $b_0$ and mass $\mu$ related 
by \cite{footnote2} 
\be \label{bmu}
\mu = -\frac{\pi b_0}{2} \;.
\ee
Gonzalez-Diaz extrapolates eq.~(\ref{bmu}) to a time-dependent wormhole 
embedded in a FLRW universe and proceeds to estimate the accretion rate on 
such a wormhole by adopting the formula derived for stationary accretion 
onto a black hole embedded in a FLRW universe \cite{Babichevetal}
\be \label{Babichev}
\dot{\mu}=4 \pi D  \mu^2  \left( P+\rho \right) \;,
\ee 
where $D$ is a constant.  By 
combining eqs.~(\ref{bmu}) and (\ref{Babichev}), Gonzalez-Diaz obtains an 
equation for the rate of change of the wormhole throat $\dot{b}_0$ and 
proceeds to solve it in a FLRW universe approaching the Big Rip, 
concluding that the wormhole shell of radius $ b_0(t)$ expands faster than 
the universe and 
ends up engulfing it and destroying global hyperbolicity. Although 
treating the wormhole as a black hole 
with negative mass seems reasonable, unfortunately eq.~(\ref{bmu}) does 
not  hold for a time-dependent wormhole embedded in a FLRW universe -- cf. 
eq.~(\ref{sigma}) or eq.~(\ref{M}) for the first wormhole model, or 
eq.~(\ref{label}) for the second wormhole model presented. The 
time  evolution of the wormhole shell cannot be guessed {\em a priori}  
but needs to be derived from detailed models like the ones presented 
here. 

The first wormhole model studied in this paper is 
appropriate for elucidating the physical situation considered in 
Ref.~\cite{Gonzalez}. In fact, by comparing  eq.~(2) 
of Ref.~\cite{Gonzalez} and our eq.~(\ref{accretion}), one  
sees that the regime of stationary accretion onto the wormhole 
considered in Ref.~\cite{Gonzalez} corresponds to the situation 
$v=$constant in our formalism, for which the dynamics of the shell has 
been explicitly solved (the static limit of such a solution 
would correspond to $a \equiv 1$ and $ \dot{M}=-8\pi P_{(\Sigma)} 
R\dot{R}$). In 
this case the wormhole shell ends up expanding at the same rate as the 
universe in which it is embedded.
However, this  wormhole model has one 
limitation, namely the assumption that  the wormhole does not perturb the 
surrounding FLRW universe.  More general exact solutions describing a 
wormhole embedded in, and modifying,  the surrounding cosmological 
background are presented in Sec.~5. This more general  class of exact 
solutions contains three free functions 
$a(t)$, $m(t)$, and $ R(t)$. By imposing the form (\ref{bigrip}) of the 
scale factor 
$a(t)$ appropriate to the description of a Big Rip and by 
noting that the metric component $g_{00}=-\left(B/A \right)^2 $ is bounded 
even if $m(t)$ diverges, one deduces again that the  
wormhole shell becomes comoving with the cosmological background as the  
Big Rip is approached.

\section*{Acknowledgments}

This work was supported by the Natural Sciences  and Engineering Research 
Council of Canada ({\em NSERC}).


   
\end{document}